# MALTS: A tool to simulate Lorentz Transmission Electron Microscopy from micromagnetic simulations

Stephanie K. Walton[1], Katharina Zeissler[1], Will R. Branford[1,2], and Solveig Felton[2,3]

[1]Department of Physics, [2]London Centre for Nanotechnology, [3]Department of Materials,
Imperial College, London SW7 2AZ, UK

**Here we describe the development of the MALTS software which is a generalised tool that simulates Lorentz Transmission Electron Microscopy (LTEM) contrast of magnetic nanostructures. Complex magnetic nanostructures typically have multiple stable domain structures. MALTS works in conjunction with the open access micromagnetic software Object Oriented Micromagnetic Framework or MuMax. Magnetically stable trial magnetisation states of the object of interest are input into MALTS and simulated LTEM images are output. MALTS computes the magnetic and electric phases accrued by the transmitted electrons via the Aharonov-Bohm expressions. Transfer and envelope functions are used to simulate the progression of the electron wave through the microscope lenses. The final contrast image due to these effects is determined by Fourier Optics. Similar approaches have been used previously for simulations of specific cases of LTEM contrast. The novelty here is the integration with micromagnetic codes via a simple user interface enabling the computation of the contrast from any structure. The output from MALTS is in good agreement with both experimental data and published LTEM simulations. A widely-available generalized code for the analysis of Lorentz contrast addresses is a much needed step towards the use of LTEM as a standardized laboratory technique.**

## I. INTRODUCTION

THE SENSITIVITY of electrons to local magnetic fields enables Lorentz Transmission Electron Microscopy (LTEM) to probe magnetic microstructure [1]-[16]. The LTEM contrast patterns which result from simple magnetic structures are well established, with the borders between domains showing up as bright or dark lines in the Fresnel or defocus mode [1]-[3]. However for more complex magnetic structures, the LTEM contrast is not so intuitive or easy to understand making simulation important. Several different groups [5]-[12], [16], [17] have published simulations of LTEM contrast, obtained using code based on similar equations, but not generally available, rendering comparison between different simulations challenging. For example Qi et al. [5], [6] use a simple MATLAB program which works on structures uniformly magnetised in the x-direction, while McVitie and Cushley [7] have a more complex simulator capable of studying multiple domain structures. We have developed MALTS (Micromagnetic Analysis to Lorentz TEM Simulation) to serve as a transparent and easy-to-use software that computes Fresnel mode LTEM contrast images for thin magnetic nanostructures of all complexities.

The publicly available Object Oriented Micromagnetic Framework (OOMMF) [18] and MuMax [19] software enable the groundstate of magnetic structures to be computed as a function of applied magnetic field. This is achieved by solving the Landau-Lifshitz-Gilbert equation in defined meshes using numerical integration. The result can be saved as an .omf file and displayed either showing the magnetisation direction of each mesh allowing comparison with X-ray Magnetic Circular Dichroism images, or showing the divergence of the magnetisation, useful for comparison with Magnetic Force Microscopy images. However, a representation of a corresponding Lorentz TEM image is not available. Here we describe how MALTS can convert the outputted .omf file from OOMMF or MuMax into a corresponding LTEM image. Similar software called GENIUS was presented by Haug et al. [17] in 2003, but we were unable to obtain it. We provide our MALTS both as precompiled executables and as open source code, allowing users to expand and improve on the functionality.

## II. METHOD

In Lorentz TEM, some of the incident high energy electrons are transmitted through the sample. These electrons experience a Lorentz force due to both local magnetic and electric components. These interactions can be expressed in terms of a phase via the Aharonov-Bohm expression [20]

$$\varphi = \varphi_e + \varphi_m = C_E \int V dz - \frac{\pi}{\Phi_0} \int A_z dz \tag{1}$$

where $\varphi_e$ is the electric phase, $\varphi_m$ is the magnetic phase, $C_E$ is the accelerating voltage constant, $\Phi_0$ is the magnetic flux quantum, $V$ is the inner potential of the material and $A_z$ is the z-component of the magnetic vector potential, where the axes are defined in Fig. 1.

**FIG. 1 HERE.**

The electric phase term can be rewritten as $\varphi_e = C_E V_0 t$, where $V_0$ is the mean inner potential of the material and $t$ is the thickness. The magnetic term, however, is more complex. Assuming that the x- and y-components of magnetisation vary only with the x- and y-coordinates, the magnetic phase component can be simplified in reciprocal space [21] to

$$\widetilde{\varphi}_m = \frac{i\pi\mu_0 M_S t}{\Phi_0}\left(\frac{\widetilde{m}_x k_y - \widetilde{m}_y k_x}{k_x^2 + k_y^2}\right) \tag{2}$$

where $\mu_0$ is the permeability of free space, $M_S$ is the saturation magnetisation of the material, $\widetilde{m}_x$ and $\widetilde{m}_y$ are the magnetisation unit vectors in reciprocal space, and $k_x$ and $k_y$ are the x- and y-components of the reciprocal space k vector. This assumption holds over a single mesh in the z direction. MALTS deals with multiple meshes in the z direction via the linear addition of magnetic phases accrued through each individual mesh. For most TEM specimens, however, a single mesh in the z direction is a reasonable assumption since the film thickness is generally much smaller than the lateral dimensions. As such, all MALTS simulations demonstrated here have a single mesh in the z direction.

The sample may be tilted in order to detect out-of-plane magnetisation; in experimental LTEM a sample tilt may also be used to apply an in-plane magnetic field. If the sample is tilted $\beta$ degrees about the x-axis, the magnetisation unit vectors must be computed in a different coordinate system via $m_y = m'_y \cos\beta - m'_z \sin\beta$ and $m_x = m'_x$, see Fig. 1(a).

Proceeding in this manner, this can be generalised to tilt of $\beta$ about an arbitrary axis in the xy-plane, $\theta$ degrees measured from the x-axis towards the y-axis, see Fig. 1(b):

$$\begin{aligned}m_x &= m'_x(\cos^2\theta + \sin^2\theta\cos\beta)\\&+ m'_y(\sin\theta\cos\theta)(1-\cos\beta) + m'_z\sin\beta\sin\theta\end{aligned} \tag{3}$$

$$\begin{aligned}m_y &= m'_x(\sin\theta\cos\theta)(1-\cos\beta)\\&+ m'_y(\cos\beta\cos^2\theta + \sin^2\theta) - m'_z\sin\beta\cos\theta\end{aligned} \tag{4}$$

In addition, the sample has a new effective thickness due to the tilt, $t = t'/\cos\beta$, and its new projection on the xy-plane is accounted for by resizing the sample via a bicubic interpolation method. Once the Fourier transform of the reciprocal magnetic phase shift has been calculated, the two phase terms can be added linearly, resulting in a net phase.

When the electrons have passed through the structure acquiring both a magnetic and an electric phase, they reach the back focal plane of the objective lens. Here the electron disturbance can be computed by performing a Fourier transform on the wave function of the transmitted electron beam.

$$g(k_x, k_y) = \iint f(x, y)\exp(-2\pi i(k_x x + k_y y))dxdy \tag{5}$$

Since all electron lenses are finite in size and are subject to aberrations, the electron wave is modified to $g(k_x, k_y)t(k_x, k_y)$ by the "transfer function" [1]:

$$t(k_x, k_y) = A(k_x, k_y)$$
$$\exp\left\{-2\pi i\left(\left[\frac{C_S\lambda^3(k_x^2 + k_y^2)}{4}\right] - \left[\frac{\Delta z\lambda(k_x^2 + k_y^2)}{2}\right]\right)\right\} \tag{6}$$

in which $\lambda$ is the relativistic wavelength of the electrons. This modification depends on both the spherical aberration coefficient of the (effective) objective lens, $C_s$, and the defocus, $\Delta z$. In MALTS the pupil function $A(k_x, k_y)$ is assumed to be constant for all reciprocal space. Since Fresnel mode LTEM involves using a large defocus the term involving the spherical aberration is small compared to the defocus term and usually has a negligible effect, so most of our simulations were performed at $C_s = 0$. However, since spherical aberration varies from instrument to instrument the user is able to input their instrument's spherical aberration for simulations.

For a real microscope the fact that the resolution is limited by the spatial coherence and spread of the electron source needs also be taken into account. MALTS uses an envelope function describing the spread of the source as a Gaussian distribution [22]:

$$E_S(k) = \exp\left[-\left(\frac{\pi\alpha}{\lambda}\right)^2(C_S\lambda^3 k^3 + \Delta z\lambda k)^2\right] \tag{7}$$

in which $\alpha$ is the beam divergence angle and $k = (k_x^2 + k_y^2)^{1/2}$. The envelope function acts to dampen the electron signal at high scattered angles. Finally an inverse Fourier transform is required to get the final intensity at the screen.

$$I(x', y') =$$
$$\left|\iint g(k_x, k_y)t(k_x, k_y)E_S(k)\exp[-2\pi i(k_x x' + k_y y')]dk_x dk_y\right|^2 \tag{8}$$

The input file into MALTS is an .omf file specifying the x, y, and z magnetisation components at each mesh. Since Fourier Optics is required and Discrete Fourier Transforms are best performed on vectors of size $2^N$ where N is an integer, it is necessary to then "zero pad" the magnetisation matrix to a larger matrix of size $2^N$. The user is able to decide the matrix size, provided that it exceeds the size of the inputted file, and hence they can dictate the amount of zero padding. Zero padding is considered physical since the electrons are incident on an area far larger than the magnetic structure actually occupies. Larger amounts of zero padding are therefore in general more similar to the actual Lorentz TEM situation. Increasing the amount of zero padding leads to increased computational time, but even for the largest matrix size available in MALTS, 2048, the entire image simulation process takes less than a minute.

## III. VALIDATION

In order to test MALTS, comparisons have been made with both experimental LTEM images and published LTEM simulations from other groups [5]-[7], [11], [12].

Fig. 2 shows the LTEM simulations from MALTS for exactly the same dimensions specified in Qi's thesis [6] and displayed in figure 3 of Qi et al [5], i.e. a bar of Permalloy, 512 nm long, 100 nm wide, and 22 nm thick, uniformly magnetised along its long axis as shown in Figs. 2(a) and 2(b).

**FIG. 2 HERE.**

Three different simulations are performed with different amounts of zero-padding of the matrix: i) No zero-padding where the magnetisation extends to the left and right hand edge of the matrix (Fig. 2(c) and (d)). ii) Zero-padding to make the matrix twice as wide as the magnetic pattern (Fig. 2(e) and (f)). iii) Zero-padding to make four times as wide as the element (Fig. 2(g) and (h)). The zero-padding case in Figs 2(c) and (d) corresponds to the simulations performed by Qi et al. [5], [6]. The MALTS simulation without zero-padding shows good agreement with Qi et al.'s [5] simulation. It is striking that the inclusion of zero-padding, such that the magnetic structure of interest is clear of the image edges, significantly changes the simulated images 2(e) and (f), compared to 2(c) and (d). Once the magnetic structure is well within the image boundaries (at twice the largest dimension of the magnetic element), further zero-padding does not significantly alter the simulated images in Figs. 2(g) and (h).

All three cases exhibit the principal contrast feature of bright contrast on the upper and lower parts of the bar when the bar is magnetised to the left and right respectively, which is sufficient to correctly attribute the LTEM image in such a simple case. However in the analysis of more complex structures it is important to place the features of interest well away from the edges of the matrix to avoid confusing real contrast with edge effects arising due to the assumption of periodicity in the fast Fourier transform.

MALTS was also used to obtain LTEM simulations of four domain flux closure states in a 1 µm × 2 µm rectangular element of 20 nm thickness. Four defocus values − 5 µm, 100 µm, 1500 µm and 10000 µm − were chosen to facilitate comparison between MALTS (see Fig. 3) and the simulations of McVitie and Cushley's figure 9 [7].

**FIG. 3 HERE.**

For defocus values of 5 µm (Figs 3 (c) and (d)) and 100 µm (Figs. 3 (e) and (f)) the MALTS simulations are in excellent agreement with McVitie and Cushley's [7]: filamentary bright or dark fringes mark the borders between domains of differently oriented magnetisation for the cases of clockwise or anticlockwise rotation of the magnetisation respectively. As expected for Fresnel mode LTEM, inverting the sign of the magnetisation changes bright lines to dark and vice versa. This agreement between MALTS and McVitie and Cushley's simulations [7] does not extend to the largest defocus of 10000 µm (Figs. 3 (i) and (j)). However similar contrast between our simulation at a defocus of 1500 µm (Figs. 3(g) and (h)) and theirs at 10000 µm was seen. The disparity at this very large value of defocus could be due to the use of different values for the beam divergence; McVitie and Cushley [7] do not state what value they use. Another possible reason for this discrepancy is the use of different approximations in the respective software, e.g. the envelope function used which again McVitie and Cushley [7] do not specify.

Phatak et al. [11] reported that tilting the sample enabled the study of a vortex core's polarity in nanodiscs, i.e. the direction of the out-of-plane magnetisation, something the electrons would otherwise be insensitive to. Simulations were carried out with MALTS under similar conditions, excluding the introduction of local magnetocrystalline anisotropy in the vicinity of the core.

**FIG. 4 HERE.**

Fig. 4(a-d) show the contrast obtained from anticlockwise chirality up polarity, anticlockwise chirality down polarity, clockwise chirality up polarity and clockwise chirality down polarity respectively at a tilt of 34° about the x-axis. (Up and down polarity are defined as being in the positive and negative z-direction respectively, see Fig. 1.) These images have a dark or bright core for anticlockwise or clockwise chirality, in agreement with the results of Phatak et al. [11]. Differences in the contrast can be seen for the same chirality but different polarity configurations. The plots of the intensity variation along a line through the core shown in Figs. 4(e) and (f) show that the polarity affects both the position of the core and the profile of the intensity peak. The latter effect was also observed by Phatak et al. [11]. However, they do not mention any shift of the core and it is difficult to tell from their figures whether their simulations also produced this effect [11]. However Ngo and McVitie [12] illustrated a new approach to determining the core polarity in nanodiscs, albeit of slightly different dimensions (600nm diameter and 20nm thick) to Phatak et al. They suggested that, by subtracting the contrast of an LTEM image taken at negative tilt from one taken at positive tilt, the core's polarity could easily be ascertained: this created a white-and-black spot where the position of the white and the black contrast depend on the polarity. MALTS simulations in Fig. 5 support this methodology; Figs. 5(a) and (b) show the difference images for nanodiscs of the same anticlockwise chirality but different polarity ((a), up, (b), down) between +30° and -30° tilt, clearly demonstrating the inversion of the black-and-white contrast for different polarity cores. The relative position of the black and white spots for a given polarity is reversed for the MALTS simulation compared to Ngo and McVitie [12]. We assume that this is due to a different assignment of the positive tilt direction. Fig. 1 shows our definition of positive tilt direction; Ngo and McVitie [12] do not explicitly define theirs.

**FIG. 5 HERE.**

Figs. 5(c) and (d) show the intensity profiles across the core in the situations of no tilt, +30° tilt and -30° tilt, as well as the difference between the latter two tilts.

Comparison of MALTS simulations with our own experimental Fresnel mode LTEM images of two more complex nanostructures was also carried out.

**FIG. 6 HERE.**

The first structure is a set of five nanobars (100 nm × 1000 nm) relaxed in a saturating field in the negative x-direction, in a double-Y shaped geometry (Fig. 6(a)) and the second is a cross structure consisting of four 1 μm × 3 μm rectangular elements connected by 100 nm wide lines (Fig. 6(b)). MALTS simulations are shown in Figs. 6 (c) and (d). Both structures were manufactured using e-beam lithography, thermal evaporation of a 20 nm thick Permalloy ($Ni_{81}Fe_{19}$) layer, and lift-off on a 50 nm thick $Si_3N_4$ membrane for TEM from Agar Scientific (Figs. 6(e) and (f)). A 5 nm layer of Au was sputtered onto the sample to avoid charge build up under the electron beam. Good agreement between simulation and experiment was achieved in both cases, although the simulation images were sharper. This may be explained by the fact that the simulation only takes into account the magnetic Permalloy layer, while in the experimental case further scattering of the electron beam may take place in the $Si_3N_4$ membrane and the Au film.

The five nanobar structure (Fig. 6(c)) showcases the ability of MALTS to reproduce single domain contrast, such as that simulated by Qi et al. [5], [6] in more complex structures. The cross structure (Fig. 6(d)) takes this one step further demonstrating that MALTS can simultaneously produce the traditional domain-boundary contrast associated with Fresnel mode LTEM as well as single domain contrast, in relatively large structures.

## IV. HOW TO USE MALTS

MALTS is a standalone executable which is used in conjunction with the OOMMF [18] or MuMax [19] software, also publicly available. MALTS is available as a 32 or 64 bit compiled version at http://www3.imperial.ac.uk/people/w.branford/research with an accompanying user manual and MCR installer, as well as MATLAB source code. Supplying MALTS as open source enables users to extend the functionality of the software, including adding other imaging modes such as Foucault should this be desired. MATLAB was chosen as the programming language because it is designed for matrix manipulation and has inbuilt graphing functions.

MALTS requires one input text file from OOMMF or MuMax as well as the user defined values used to compute this file: material thickness, mesh size and number of meshes. In addition experimental values, "beam divergence", "defocus", "spherical aberration", and "accelerating voltage" specific to individual experiments may be varied. The user can also choose the size of the calculation matrix and thereby the amount of zero padding of the magnetic structure. The sample may be tilted in the simulation about any axis in the xy-plane, see Fig. 1. The resulting LTEM contrast is displayed on the Graphical User Interface and saved automatically. To aid in determining the origin of the LTEM contrast, images can also be simulated using only the electric or only the magnetic phase by selecting the "Electric Component LTEM" button or by setting the mean inner potential to zero respectively.

## V. CONCLUSION

In conclusion, MALTS provides a generic platform for the effective analysis of Fresnel contrast in Lorentz TEM images of magnetic structures of arbitrary shape. This will enable reproducible analysis of LTEM images and direct comparison of results across groups making LTEM more accessible to non-specialist users.

## ACKNOWLEDGMENT

The authors would like to thank Hak-Sung Lee, Scott Findlay and David Ellis for useful discussions. The authors gratefully acknowledge funding from the EPSRC (EP/D063329/1), EPSRC Career Acceleration Fellowships (EP/G004765/1) and the Leverhulme Trust (F/07058/AW).

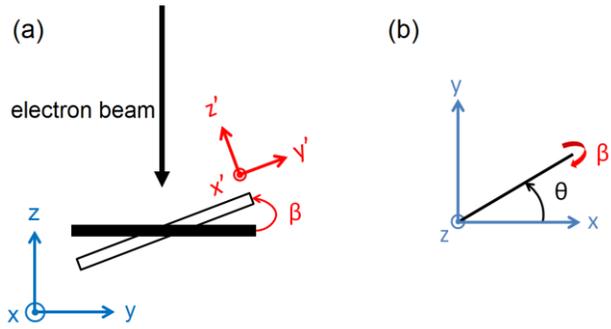

Fig. 1. Schematic showing the coordinate systems used, x,y,z for the electron beam in blue and x',y',z' for the sample in red. The angles β and θ described in the text are also illustrated. β is the angle of tilt towards the incoming electron beam, as shown in (a) in which the solid rectangle represents the sample perpendicular to the beam and the unfilled rectangle shows the sample tilted an angle β. θ defines the axis in the xy-plane about which this rotation is performed as shown in (b).

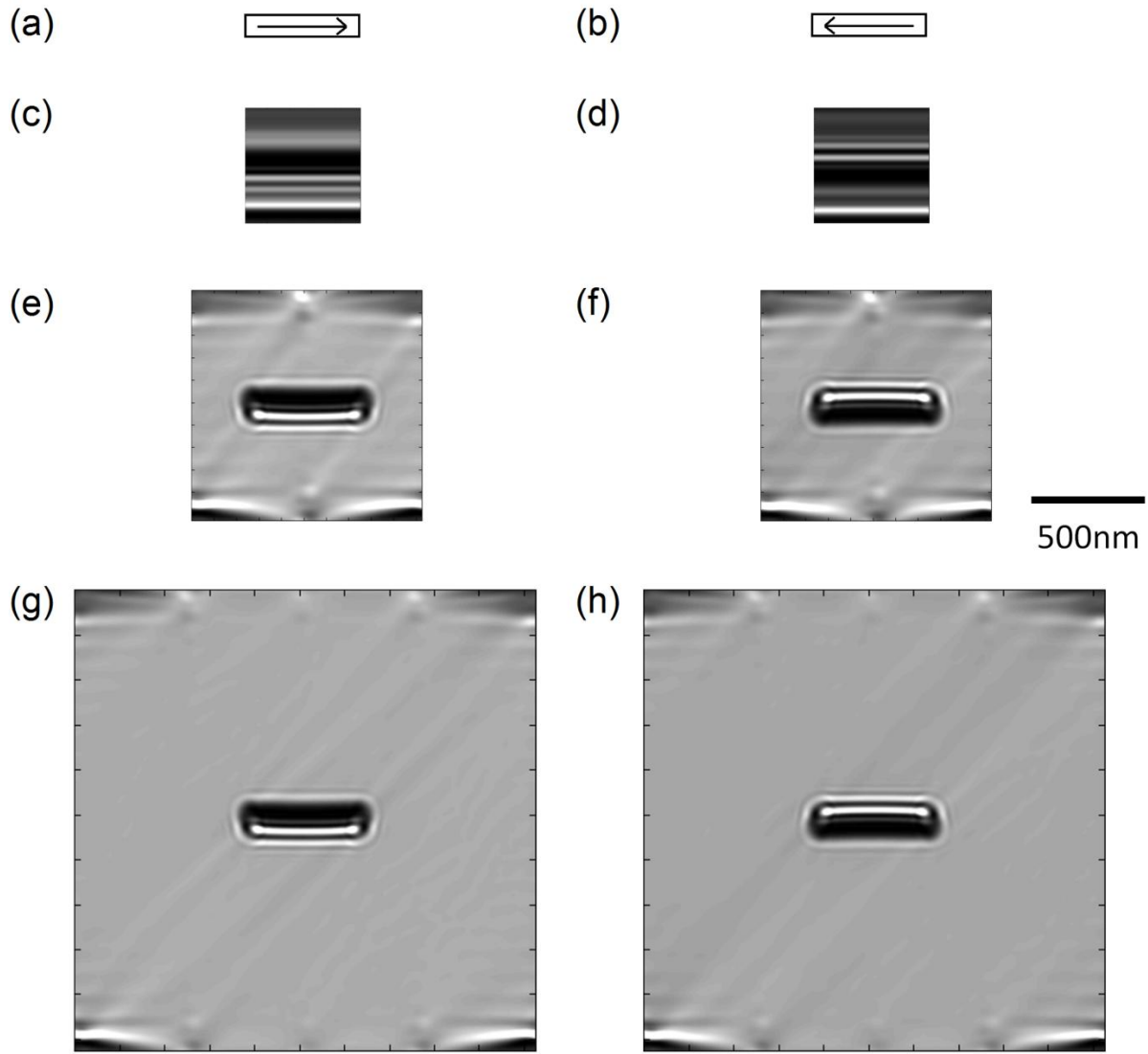

Fig. 2. MALTS simulations of Lorentz TEM images showing the effects of zero padding for a uniformly magnetised Permalloy bar of thickness 22 nm, lateral dimensions 512 nm × 100 nm and a mesh size of 1nm as specified in the thesis by Qi [6]. (a) and (b) show the magnetisation directions of the bars used for the simulations. (c) and (d) show the corresponding LTEM simulation for a matrix size of 512 × 512, (e) and (f) for matrix size 1024 × 1024 and (g) and (h) for 2048 × 2048. When the bar is magnetised in the right (left) direction, bright contrast is seen on the lower (upper) side of the bar. An accelerating voltage of 300 kV, a defocus of 1600 μm, a spherical aberration of 0 m, and a beam divergence of 0.01 mradians were used for the simulations.

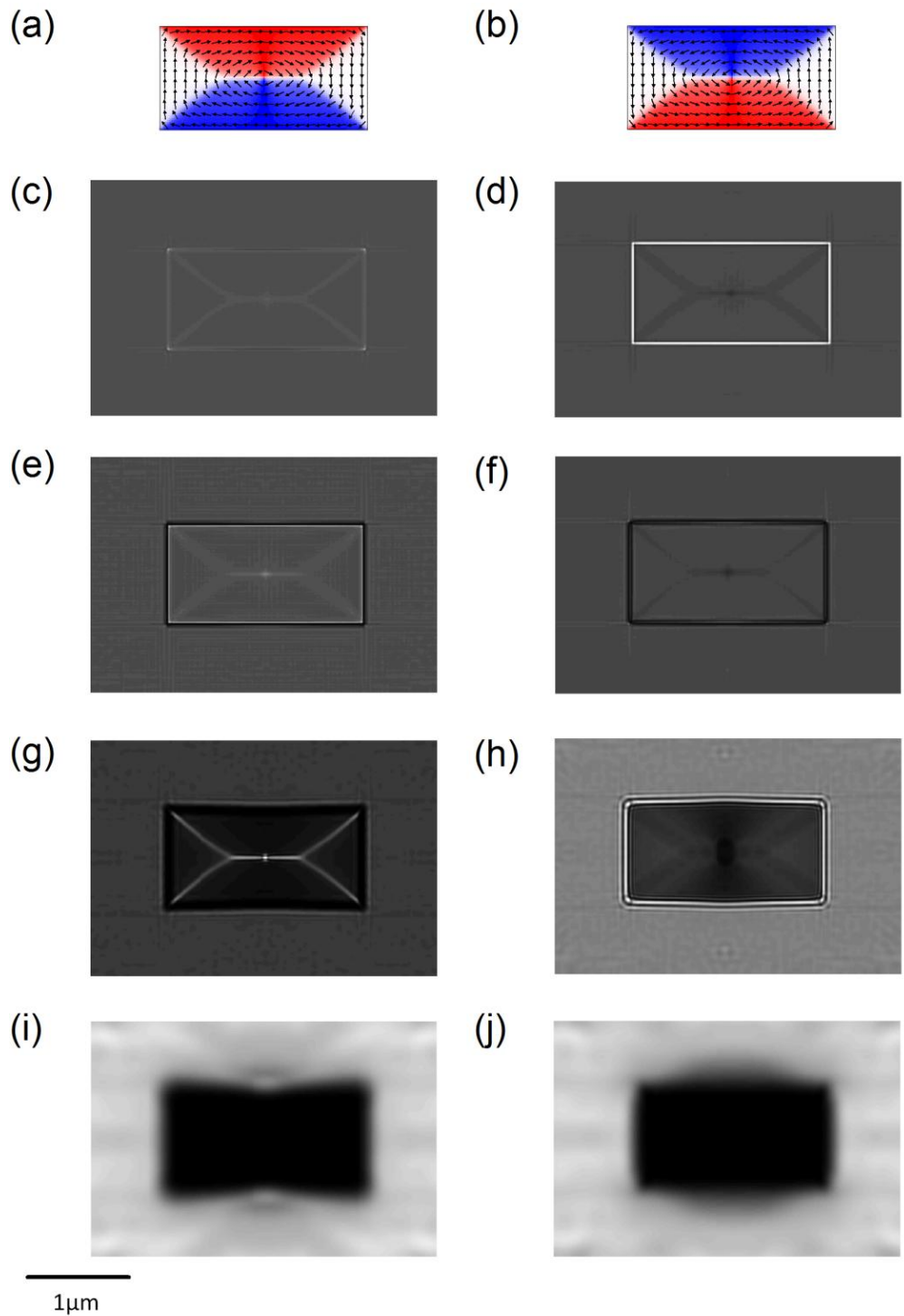

Fig. 3. Micromagnetic simulation of clockwise (a) and anticlockwise (b) four-domain flux-closure states in rectangular 1 μm × 2 μm, 20 nm thick Permalloy nanostructures, corresponding to the dimensions used by McVitie and Cushley [7]. The red and blue colours indicate magnetisation in the positive and negative x-directions respectively. (c)-(j) MALTS LTEM simulations for the magnetic state in (a) and (b) in the left and right hand columns respectively. For comparison with figure 9 of reference [7], images at defocus (c) and (d) 5 μm, (e) and (f) 100 μm, (g) and (h) 1500 μm and (i) and (j) 10000 μm were produced. An accelerating voltage of 200 kV, a $C_s$ of 8000 mm, and a beam divergence of 0.01 mradians were used. The mesh size was 5 nm and the matrix size used for zero padding was 1024.

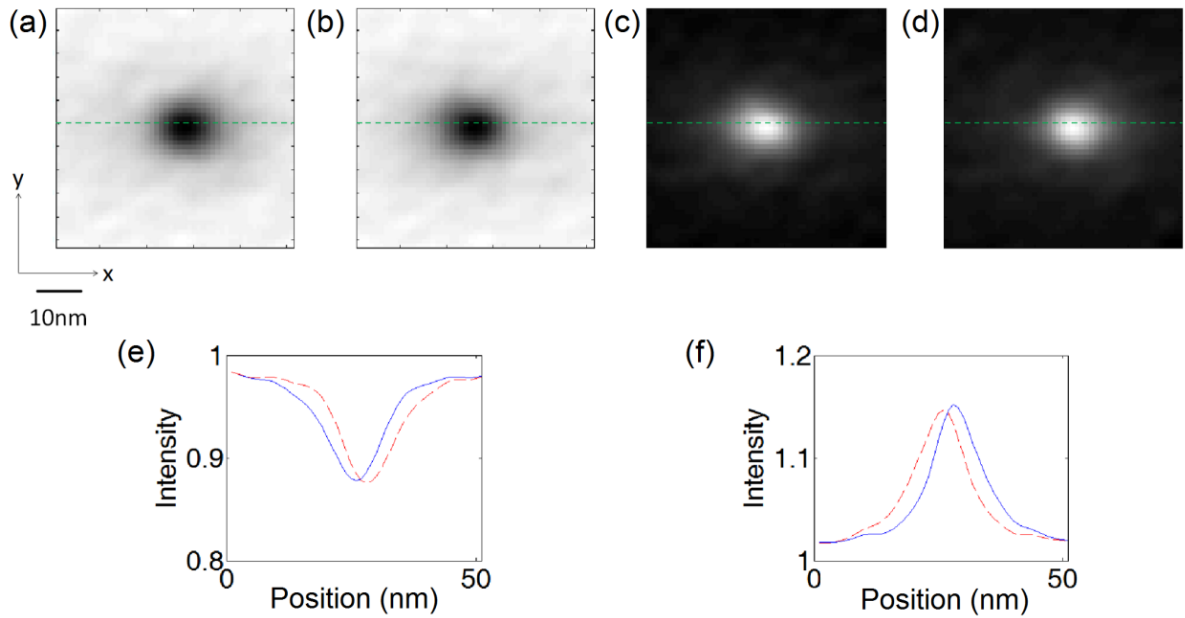

Fig. 4. MALTS simulations of LTEM images for 15 nm thick nanodisc of 250 nm radius with exchange constant A = 30.2 pJ/m and $M_s$ = 2 T and mesh size 1 nm, matching the parameters for similar simulations performed by Phatak et al. [11]. An accelerating voltage of 200 kV, a $C_s$ of 1 m, a beam divergence of 0.01 mradians and a defocus of 5 μm were used. The matrix size was 1024 × 1024. (a), (b), (c) and (d) show the contrast obtained at a 34° tilt about the x-axis when an anticlockwise vortex with up polarity, an anticlockwise vortex with down polarity, a clockwise vortex with up polarity and a clockwise vortex with down polarity respectively are simulated. The green dotted line indicates where the cross sections of intensity, (e) and (f) have been taken. The red dashed and blue solid lines in (e) and (f) indicate the cross sections through (a) and (b) in (e), and (c) and (d) in (f) respectively.

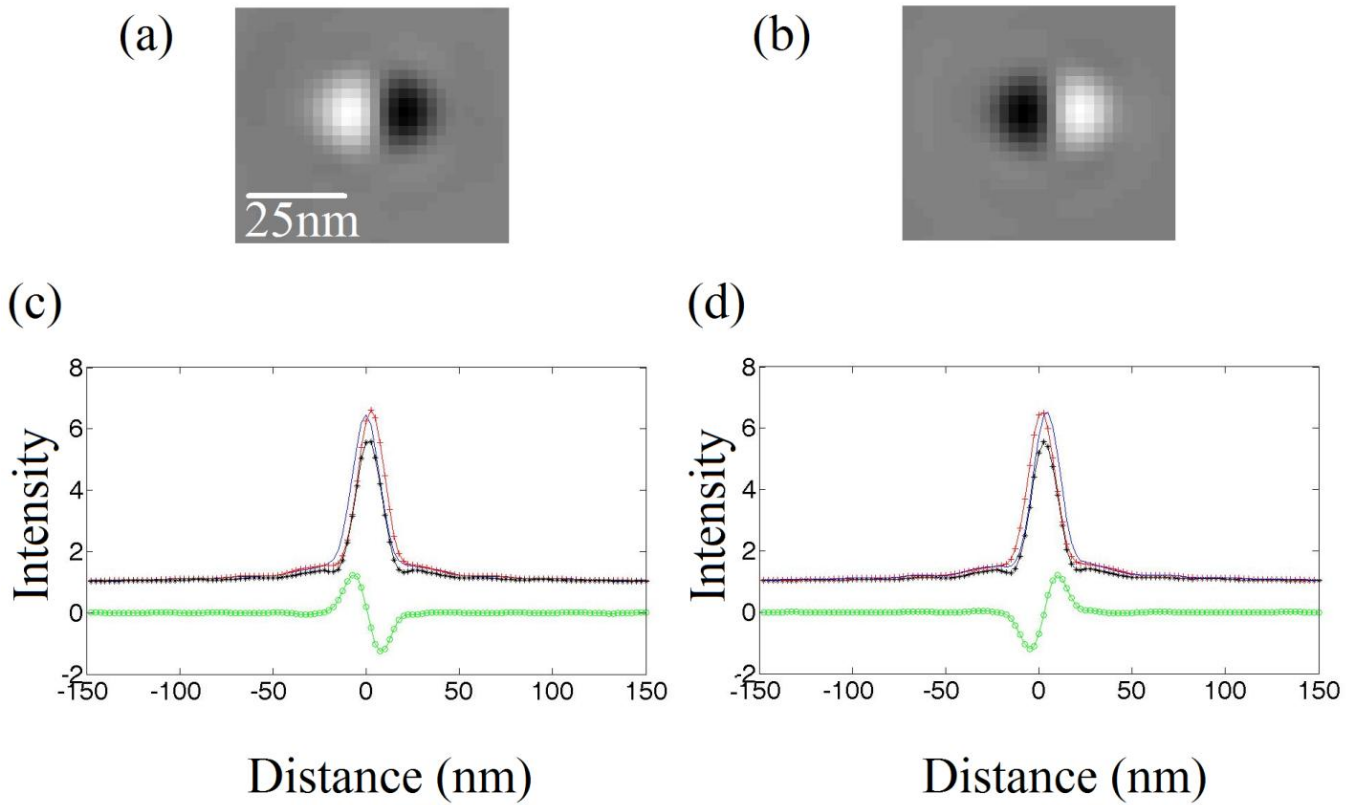

Fig. 5. MALTS simulations of LTEM images for 20 nm thick nanodisc of 600 nm diameter with exchange constant A = 13 pJ/m and $M_s$ = 1 T and mesh size 2.5 nm, matching the parameters for similar simulations performed by Ngo and McVitie [12]. An accelerating voltage of 200 kV, a $C_s$ of 8000 mm, a beam divergence of 0.01 mradians and a defocus of 250 μm were used. The matrix size was 1024 × 1024. (a) and (b) show the difference contrast obtained between a +30° at at -30° tilt about the x-axis when (a) an anticlockwise vortex with up polarity and (b) an anticlockwise vortex with down polarity are simulated. (c) and (d) show the intensity profiles across the core for an anticlockwise vortex with up polarity and clockwise vortex with down polarity respectively. The black asterisked line shows the intensity profile at zero tilt. The red crossed line and the blue line show the profiles at +30° and at -30° tilt respectively. The green line with open circles shows the difference between the intensity profile at +30° and at -30° tilt. The relative position of the black and white spots for a given polarity is inversed in these simulations compared to those of Ngo and McVitie [12], which we assume to be due to a different definition of positive tilt direction.

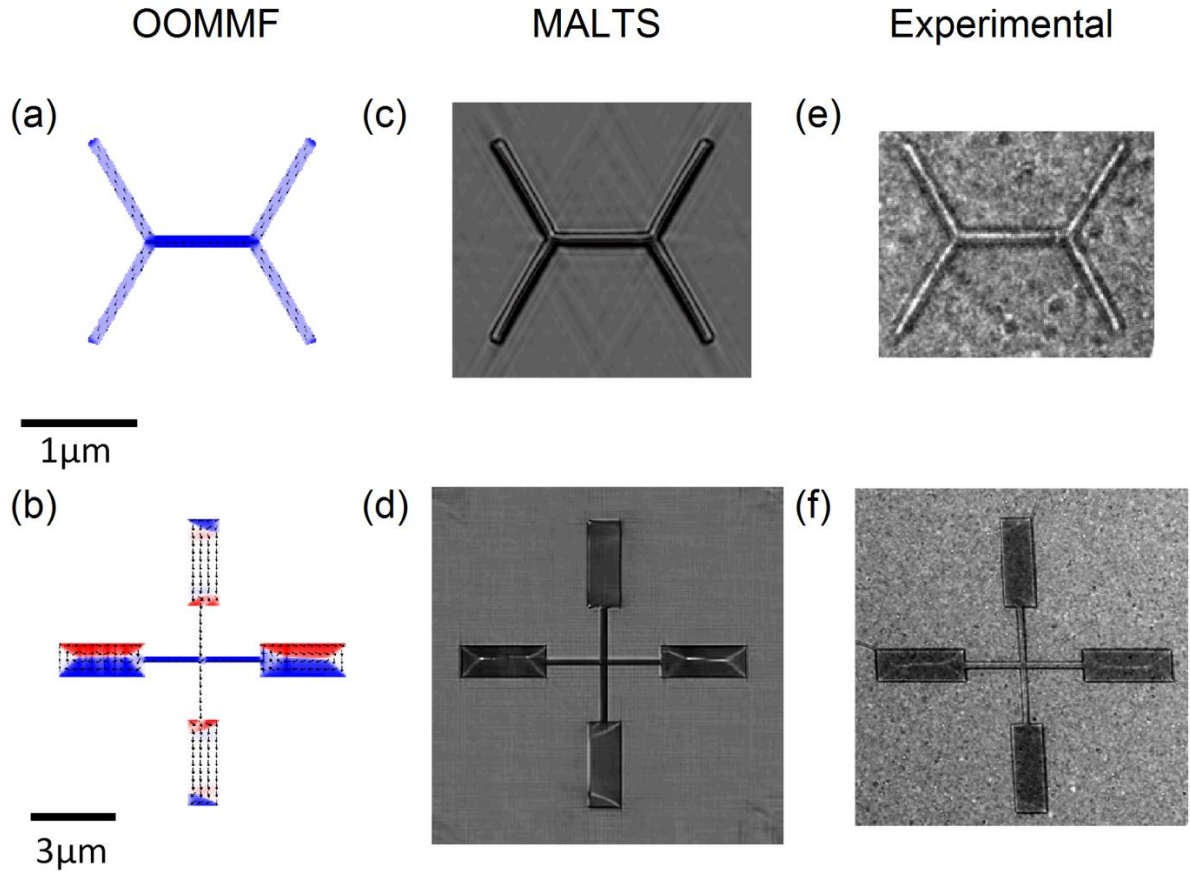

|      | OOMMF | MALTS | Experimental |
|------|-------|-------|--------------|

Fig. 6. Comparison of experimental data with LTEM images simulated by MALTS for 20 nm thick Permalloy nanostructures. The micromagnetic states of (a) five bars saturated in the negative x-direction and (b) a cross structure with both single domain and flux closure behaviour, were simulated using OOMMF. The red and blue colours indicate magnetisation in the positive and negative x-directions respectively. (c, d) LTEM simulations using an accelerating voltage of 300 kV, a defocus of 1.5 mm, a $C_s$ of 0 m, and a beam divergence of 0.01 mradians. (e, f) Experimental LTEM images obtained using an accelerating voltage of 300 kV and a defocus of 1.5 mm. Mesh sizes of 5 nm and 10 nm and matrix sizes of 1024 and 2048 were used for the five bars and the cross structure, respectively.